\documentclass[english,floatfix,twocolumn,nobalancelastpage,superscriptaddress,nofootinbib,prd]{revtex4}
\usepackage[paperwidth=210mm,paperheight=297mm,centering,hmargin=2cm,vmargin=2cm]{geometry}
\usepackage[utf8]{inputenc}
\usepackage[T1]{fontenc}
\usepackage{lmodern}
\usepackage{amsmath}
\usepackage{amssymb}
\usepackage{graphicx}
\usepackage{esint}
\usepackage{longtable}
\usepackage{dcolumn}
\usepackage{babel}
\usepackage{xcolor} 
\usepackage{csquotes}
\usepackage{marvosym}
\usepackage{hyperref}
\usepackage[overload]{textcase}
\usepackage[symbol,hang,flushmargin]{footmisc}

\DefineFNsymbolsTM{otherfnsymbols}{%
  \Radioactivity	\circ
  \Yinyang   \mathsection
  \Bat    \dagger
}%

\setfnsymbol{otherfnsymbols}

\newenvironment{seqn}{\equation\aligned}{\endaligned\endequation}

\newcommand{\be}{\begin{seqn}}
\newcommand{\ee}{\end{seqn}}

\newcommand{\bea}{\begin{eqnarray}}
\newcommand{\eea}{\end{eqnarray}}

\newenvironment{arabicfootnotes}
  {\par\edef\savedfootnotenumber{\number\value{footnote}}
   
   \setcounter{footnote}{0}}
  {\par\setcounter{footnote}{\savedfootnotenumber}}

\begin{document}
%
%
%
%
%
%
\title{Massless Charged Particles, Naked Singularity, and Generalized Uncertainty Principle in Reissner-Nordstr\"{o}m-de Sitter-like spacetime}

\author{Elias~C.~Vagenas}
\email{elias.vagenas@ku.edu.kw}
\affiliation{Theoretical Physics Group, Department of Physics, Kuwait University, P.O. Box 5969, Safat 13060, Kuwait.}

\author{Ahmed~Farag~Ali}
\email{ahmed@quantumgravityresearch.org} 
\affiliation{Department of Physics, Faculty of Science, Benha University, Benha, 13518, Egypt.}
\affiliation{Quantum Gravity Research, Los Angeles, CA 90290, USA}

\author{Hassan~Alshal}
\email{halshal@sci.cu.edu.eg}
\affiliation{Department of Physics, Faculty of Science, Cairo University, Giza, 12613, Egypt}
\affiliation{Department of Physics, University of Miami, Coral Gables, FL 33146, USA}

\begin{abstract}
\vspace{0.2cm}
\section*{Dedication}
\vspace{-0.2cm}
\noindent One of us (A.F.A.) wishes to dedicate this work to the memory of his late father Farag Mohammad Ali whose sudden death during the preparation of the final draft of this work in 2019 greatly sorrowed him.\\
\vspace{-0.5cm}
\section*{Abstract}
\vspace{-0.2cm}
\par\noindent
 Motivated by the endeavors of Li  Xiang and You-Gen Shen  on naked singularities, we investigate the validity of the cosmic censorship conjecture in the context of generalized uncertainty principle. In particular, upon considering both linear and quadratic terms of momentum in the uncertainty principle, we first compute the entropy of a massless charged  black hole in de Sitter spacetime at a given modified temperature. Then, we compute the  corresponding modified cosmological radius and express the black hole electric charge in terms of this modified cosmological radius and, thus, in terms of the generalized uncertainty principle parameter. Finally, we examine whether such a system will end up being a naked singularity or it might be protected by the cosmic censorship conjecture, and how that might be related to the possible existence of massless charged particles.
\end{abstract}
\maketitle
\begin{arabicfootnotes}
%
%
%
%
\section{Introduction}
%
%
%
%
%
%
\par\noindent
General Relativity (GR) is so successful not  only  because of its accurate predictions but also its  capability of predicting its shortcomings. One of the ``apparent'' shortcomings of GR is that it predicts the existence of \emph{singularities}, spacetime geometrical points where curvature becomes infinite and laws of physics are no longer working. Einstein abhorred their occurrence \cite{Earman:1995fv}. He believed they are just made-up ramifications of spacetime symmetries and morphisms. However it turned out that they are inescapable through Penrose-Hawking singularity theorems \cite{Penrose:1964wq, Penrose:1969pc, Hawking:1969sw,Hawking:1973uf}. It is worth noting that the \emph{classical} Einstein-Maxwell equations were introduced as a deterministic unified theory. However, it was shown that the Einstein-Maxwell equations can avoid singularities through, interestingly, violating energy density condition of positivity of Penrose - Hawking singularity theorems  \cite{Dolan:1968em, Finster:1998ju}. Nevertheless, this violation has undesired consequences \cite{Gibbons:1982jg, Herzlich:1998JGP} upon considering Positive Energy Theorem \cite{Schon:1979rg, Schon:1981vd, Witten:1981mf}. So the moral is that GR, alone or combined with other classical theories, fails to deal with singularities. However, this is not truly considered a shortfall as long as the singularity is \emph{causally} separated\textemdash and hence impossible to be observed\textemdash from the rest of the universe through an event horizon, a ``doorstep'' at which timelike and spacelike coordinates  interchange their rules. That makes static uncharged black hole singularities spacelike points, meanwhile those of rotating and/or charged black holes may disrupt the causal structure of spacetime. Interestingly since Penrose-Hawking singularity theorems do not say much about geometrical locations of singularities, then existence of singularity is not necessarily accompanied within black hole structure \cite{Srinivasan:1999fw}, i.e., they might not be in need of event horizon. These singularities are called naked singularities. Since a naked singularity lacks that ``doorstep'', timelike and spacelike coordinates keep their geometrical rules unchanged around that point. This would lead to a major breakdown of foundations of spacetime geometry. Consequently, the ``realistic'' perspectives on classical theories of physics would be demolished. Being a Platonist\footnote{\noindent Hawking meant ``Platonic realist''. Like Einstein, Penrose concerns about maintaining the \emph{deterministic} predictability of GR as a local theory. This philosophical stance is based on the fact that GR is exclusively a geometric theory of Lorentzian manifolds.}\cite{Hawking:1996jh}, Penrose introduced the cosmic censorship conjecture \cite{Penrose:1969pc}:
%
%
%
%
%
\begin{displayquote}
\begin{center}
``Nature abhors a naked singularity''~.
\end{center}
\end{displayquote}
%
%
%
%
%
\par\noindent
Since the introduction of cosmic censorship conjecture, many endeavors attempted to argue in favor of\textemdash mainly by focusing on Cauchy horizon \cite{Simpson:1973ua, McNamara:1978ih, Chandrasekhar:1982ch, Krolak:1986cc, Mellor:1989ac, Mellor:1992ch, Brady:1992ch, Cai:1995ux, Wald:1997wa}\textemdash or against the weak and strong versions of cosmic censorship conjecture\textemdash mainly by focusing on gravitational dust collapse processes \cite{Yodzis:1973gha, Szekeres:1979qc, Ori:1989ps, Chrusciel:1990zx, Shapiro:1991zza, Choptuik:1992jv, Christodoulou:1994hg, Joshi:1995zi, Garfinkle:1996kt, Harada:1999jf}. None came with conclusive definitive proof whether naked singularities could or could not physically exist, with a quest whether it should be in need for changing the methodology of dealing with cosmic censorship conjecture\cite{Penrose:1999vj}. Another line of research that deals with this conjecture is to see its \emph{topological} effects together with null energy condition  \cite{Joshi:1987fn, Morris:1988tu, Grubisic:1992qn, Friedman:1993ty, Schleich:1994tm, Jacobson:1994hs, Burnett:1995um, Galloway:1996hx, Galloway:1999bp, Galloway:1999br, Helfgott:2005jn, Krasnikov:2010vw, Ringstrom:2010zz, Curiel:2014zba}. Also there is another debate about the validity of either cosmic censorship conjecture or naked singularities in modified gravity theories upon considering null energy condition \cite{Bhamra:1978ns, Martinez:2005di, Bronnikov:2007ws, Bedjaoui:2010nh, Curtright:2012uz, Charmousis:2012dw, Rudra:2013mha, Rubakov:2014jja, Tretyakova:2015vaa}. Recent \emph{analytical} proof \cite{DelAguila:2018gni} is in favor of cosmic censorship conjecture, while others  \cite{Figueras:2015hkb, Crisford:2017zpi, Dafermos:2017dbw} counter-argued cosmic censorship conjecture, leaving the quest un-answered.
\par\noindent
The biggest challenge in contemporary physics is to unify GR and Quantum Mechanics (QM) in a concrete theory of quantum gravity. A little bit auspicious way, of many, to seize the ``holy grail'' is to construct a quantum field theory in curved spacetime \cite{Birrell:1982ix, Parker:2009uva}, where the curved spacetimes are black holes. This approach succeeded in introducing Hawking radiation and black hole entropy \cite{Hawking:1974rv, Hawking:1974sw} and black hole information riddle \cite{Hawking:1976ra}. Before that, Wald's \emph{classical} gedanken experiment  failed to destroy event horizon by overcharging the corresponding Kerr-Newman black hole \cite{Wald:1974ge}. If the experiment had succeeded, then\textemdash upon considering the quantum effects\textemdash the resulting naked singularity would have absorbed all black hole entropy. This would contradict the holographic  principle\cite{tHooft:1993dmi, Susskind:1994vu} as a naked singularity of Planck length size $\ell_p$ can carry only few bits of information. More recent classical gedanken experiments also support cosmic censorship conjecture upon considering different Reissner-Nordstr\"{o}m black hole \cite{Hubeny:1998ga} or upon considering the same Kerr-Newman black hole with overcharge and overspin together \cite{Sorce:2017dst}. So within the conservation of information paradigm, ``stripping'' singularity would provoke vehement information loss \cite{Mathur:2009hf}. It is worth noting that  despite the metric at singularity can be no longer \emph{regular}, i.e., it is degenerate, information may be retrieved out of the singularity even if the used technique does not work out the initial value formulation \cite{Stoica:2015fha}.
\par\noindent
So until the ``advent'' of a mature, consistent, and complete theory of quantum gravity takes place, the question of cosmic censorship conjecture remains open. But, generally, it is believed that GR and QM  ``marriage'' would happen upon some compromises. It could be necessary that QM laws need some tweaks, e.g., modifying Heisenberg Uncertainty Principle (HUP), to be compatible with a fundamental characteristic of String theory, that is energy corresponds with UV/IR increment in its length. That leads us to introduce a Generalized Uncertainty Principle (GUP) \cite{Veneziano:1986zf, Gross:1987ar, Amati:1988tn, Konishi:1989wk, Maggiore:1993rv, Garay:1994en, Scardigli:1999jh, Adler:2001vs, Das:2008kaa, Ali:2009zq} as another attempt to reconcile GR and QM.
\par\noindent
In the remainder of this work, we summarize the endeavors of Li and Shen \cite{Li:2003bs} as well as of Xiang and Shen  \cite{Xiang:2005wg} on examining the effect of the quadratic GUP on cosmic censorship conjecture. Then, we introduce both linear and quadratic GUP to  compute the GUP-modified entropy for a static spherically symmetric black hole in de Sitter spacetime. For the specific case of the massless charged Reissner-Nordstr\"{o}m-de Sitter (RN$\text{d}$S) spacetime, we find the location of the cosmological horizon  and also show that there are no more horizons so the curvature singularity is a naked singularity. Thus, one can assume that the Hawking radiation consists of massless charged particles. Therefore, we  calculate  the total energy density of those massless charged particles. Finally, the results with some concluding comments are presented.
Here,  natural units will be used, i.e., $\hbar=c=k_{B}=1$.
%
%
%
%
%
\section{Quadratic GUP effects on RN$\text{d}$S-like Spacetime}
%
%
%
%
%
%
%
\par\noindent
In this section, we summarize the analysis of Refs. \cite{Li:2003bs, Xiang:2005wg} 
starting by considering the quadratic GUP to be given by
\be \label{eq.1}
\Delta x \Delta p \geqslant 1 + \lambda \Delta p^2 
\ee
which gives an uncertainty in momentum
\be
\Delta p \sim \frac{\Delta x - \sqrt{\Delta x^2 - 4\lambda}}{2\lambda}
\label{uncertaintymom}
\ee
where $\lambda$ is the dimensionful GUP parameter which is proportional to the squared Planck length $\ell_p^2$ with $\ell_p \sim \sqrt{G}$, and  $G$ is Newton's gravitational constant. According to Planckian thermodynamics $p\sim E \sim T$. Therefore, in the presence of a quantum black hole of event horizon radius $r_h$, Eq.\eqref{uncertaintymom} can be read as
\be \label{eq.2}
T \sim \frac{r_h - \sqrt{r_h^2 - 4\lambda}}{2\lambda}~.
\ee
Similarly, for time and energy conjugates we have
\be \label{eq.3}
\Delta E \sim \frac{\Delta t - \sqrt{\Delta t^2 - 4\lambda}}{2\lambda}
\ee
where the signs in front of the radicals in Eqs. \eqref{uncertaintymom}, \eqref{eq.2}, and \eqref{eq.3} were selected  such that when we take the $\lambda \to 0$ limit,  we then obtain the conventional HUP
\be \label{eq.4}
\Delta p \sim \frac{1}{\Delta x} \qquad \text{and} \qquad \Delta E \sim \frac{1}{\Delta t}~. 
\ee
At this point, it is worth making the following comment. The GUP corrections can be assigned to the Planck constant\footnote{Here, only for this comment, we have reinstated the units of $\hbar$.}, i.e., 
$\hbar$, and thus we can define an \emph{effective} Planck constant as
\be \label{eq.5}
\hbar' \sim \hbar(1+\lambda\Delta E^2)~.
\ee
In this case, the Hawking temperature, i.e.,  $\displaystyle{T_H=\frac{\hbar}{\beta}}$, is modified and becomes
\be\label{eq.6}
T'_H \sim\frac{\hbar(1+\lambda\Delta E^2)}{\beta}
\ee
where $\beta$ is called the  \emph{reciprocal temperature} and is given as $\beta = 2\pi \kappa^{-1}$ with $\kappa$ to be the surface gravity of the black hole horizon.
\par\noindent
Since the surface gravity is inversely proportional to the black hole radius, i.e., $\kappa\sim r_h^{-1}$, the reciprocal temperature will be proportional to the black hole radius, i.e., $\beta\sim r_h$, and thus,
the temperature given in Eq. \eqref{eq.2} becomes
\bea \label{eq.7}
T'_H & =& \frac{\beta -\sqrt{\beta^2- 4\lambda}}{2\lambda}\\
& = & \frac{2}{\beta +\sqrt{\beta^2 -4 \lambda}}~.
\eea
The first law of  black hole mechanics \cite{Bardeen:1973gs} 
 for a Schwarzschild black hole in de Sitter space (SdS) \cite{Li:2003bs} reads
\be\label{eq.8}
dM = - \frac{\kappa_c}{8\pi}dA_c - \frac{V}{8\pi}d\Lambda~
\ee
where the subscript ``c'' denotes the \emph{cosmological} horizon, $\Lambda$ is the cosmological constant, and $V=\frac{4\pi}{3}r_{c}^{3}$ is the volume of  the dS universe. Now due to the problem of negative temperatures (see the factor of the differential $dA_c$ in Eq. \eqref{eq.8}),  we redefine $M$ to be the mass of everything inside the cosmological horizon, i.e., $r_c$, including the black hole mass. This defines the total energy of such a system to be $E_0=M+E_{vac}$ and we demand this total energy to be conserved. Therefore, the first law of thermodynamics 
for the cosmological horizon becomes
\be \label{eq.9}
dE_{vac}= -dM = \frac{\kappa_c}{8\pi}dA_c + \frac{V}{8\pi}d\Lambda~.
\ee
When we also demand the entropy of this system to be maximum, then the modified temperature of the black hole, i.e., $T'_H$,  becomes equal to the temperature of the cosmological horizon, i.e., $T_c$, and the entropy of cosmological horizon becomes
\be \label{eq.10}
S_c & = \int \frac{dE_{vac}}{T_c}\\
& = \frac{A_c}{4} - \int\frac{\lambda}{4\beta^2} dA_c
\ee 
where  $A_c$ is cosmological area. At this point we should point out that if we take the limit $\lambda\to 0$, we obtain  the Bekenstein-Hawking entropy of the cosmological horizon in the context of HUP. 
\par\noindent
Let us now assume we have a massless charged Reissner-Nordstr\"{o}m-de Sitter spacetime (RNdS-like) with metric
\be\label{eq.11}
ds^2 = - f(r,Q,\Lambda)dt^2+\frac{dr^2}{ f(r,Q,\Lambda)}+r^2d\Omega
\ee
with the components of the metric to be 
\be\label{metric}
 f(r,Q,\Lambda) =1+\frac{Q^2}{r^2}-\frac{\Lambda r^2}{3}
 \ee
 and the cosmological horizon, i.e., $r_c$, is defined as $f(r_c, Q,\Lambda)=0$ thus 
 the electric charge reads
 \be
Q^2=\frac{\Lambda r_c^4}{3} - r_c^2~.
\ee
The corresponding surface gravity  will be 
\be\label{eq.12}
\kappa_c = \left(\frac{2\Lambda r_c}{3} - \frac{1}{r_c}\right)~.
\ee
\par\noindent
Employing Eq. \eqref{eq.12} in order to compute the last term of Eq. \eqref{eq.10}, we get
\be\label{eq.13}
\Delta S & = -\frac{\lambda}{16\pi^2}\int \kappa^2 dA_c\\
& = - \frac{\lambda}{2\pi}\left(\ln r_c - \frac{2\Lambda}{3} r^2_c + \frac{\Lambda^2}{9} r^4_c\right)~.
\ee
Now we can consider the most probable $\Lambda$ associated with the maximum entropy by setting $\partial(\Delta S)/\partial \Lambda =0$. This will give   $r_c = \sqrt{3/\Lambda}$ which means that   the corresponding cosmological radius is equal to the radius of de Sitter spacetime that is empty of any mass and charge. 
Therefore, in the context of the quadratic GUP, the second law of black hole mechanics  forbids the existence of massless charged particles.
\par\noindent
In  Ref. \cite{Xiang:2005wg}, it was shown that, in systems such as the one under study here, 
cosmic censorship conjecture  is guaranteed by the quadratic GUP upon considering the energy-time uncertainty
\be \label{eq.14}
\Delta t \geqslant \frac{1}{\Delta E} + \lambda \Delta E
\ee
that imposes a bound on the rate of energy loss as
\be\label{eq.15}
\frac{\Delta E}{\Delta t} \sim \frac{dE}{dt}<\frac{1}{\lambda}
\ee
which in turn leads to 
\be
E\leqslant\frac{L}{\lambda}\sim\frac{1}{\sqrt{\lambda}}
\label{energy1}
\ee
where $L$ is characteristic length of the system under study. 
The reason  why Eq. \eqref{energy1}  keeps cosmic censorship conjecture safe is that for a ``covered'' black hole singularity to be a naked singularity, it is necessary the black hole to absorb the total mass of the system which means $\displaystyle{E\gg m_p\sim\frac{1}{\sqrt{\lambda}}}$. It is evident that this contradicts our result given by Eq. \eqref{energy1} in the context of quadratic GUP.
%
%
%
%
%
%
\section{Linear and Quadratic GUP Effects on RN$\text{d}$S-like Spacetime}
%
%
%
%
%
%
%
%
%
%
\par\noindent
In this section, following the previous analysis we will re-examine everything we have just summarized in \cite{Xiang:2005wg} in the light of  linear and quadratic GUP. The linear and quadratic GUP, which is also compatible with Doubly Special Relativity,  is given as \cite{Ali:2009zq} 
\be\label{eq.16}
\Delta x \Delta p \geqslant 1 - \alpha \Delta p + 4\alpha^2\Delta p^2
\ee
where if the $\alpha$-term is vanished, then if one sets $\lambda=4\alpha^{2}$, it will get Eq. \eqref{eq.1}.
The corresponding GUP-modified uncertainty in momentum is of the form
\be
\Delta p \sim & \frac{(\Delta x+\alpha) - \sqrt{(\Delta x+\alpha)^2 - 16\alpha^2}}{8\alpha^2}
\ee
and the GUP-modified uncertainty in energy will now read
\be\label{eq.17}
\Delta E \sim & \frac{(\Delta t+\alpha) - \sqrt{(\Delta t+\alpha)^2 - 16\alpha^2}}{8\alpha^2}~,
\ee
with the corresponding GUP-modified temperature to be of the form
\be\label{eq.18}
T  \sim & \frac{(r_h +\alpha) - \sqrt{(r_h +\alpha)^2 - 16\alpha^2}}{8\alpha^2}~.
\ee
As in the previous section, one can make the comment that all  linear and quadratic GUP corrections can be assigned to an {\it effective} Planck constant which is now  modified as
\be\label{eq.19}
\hbar'\sim\hbar (1-\alpha\Delta E+4\alpha^2\Delta E^2)~.
\ee
Therefore, the GUP-modified Hawking temperature becomes
\be\label{eq.20}
T'_H & = \frac{(\beta+\alpha) -\sqrt{(\beta+\alpha)^2- 16\alpha^2}}{8\alpha^2}\\
& = \frac{2}{(\beta+\alpha) +\sqrt{(\beta+\alpha)^2 -16 \alpha^2}}
\ee
which, as expected,  is  exactly the same with the one given in Eq. \eqref{eq.18}
since, as already mentioned, $\beta\sim r_h$.
\par\noindent
Now, following the analysis of the previous section, inside the cosmological horizon the modified Hawking temperature, i.e., $T'_H$,  becomes equal to the temperature of the cosmological horizon, i.e., $T_c$, the corresponding entropy of the cosmological horizon becomes
\bea \label{eq.21}
S_c \!\! & = & \!\!  \!\!  \int \frac{dE_{vac}}{T_c}\nonumber\\
& = & \!\! \!\!  \int \!\! \frac{(\beta+\alpha)+\sqrt{(\beta+\alpha)^2-16\alpha^2})}{2} \!\! \times \!\! \frac{\kappa_c dA_c}{8\pi}.
\eea
Then, we expand $S_c$ up to $\mathcal{O}(\alpha^3)$ to get
\be \label{eq.22}
S_c = \frac{A_c}{4}-\int \left(\frac{\alpha^2}{\beta^2}-\frac{\alpha}{4\beta}\right) dA_c~.
\ee
At this point, it should be stressed that this extra $\frac{\alpha}{2\beta}$ term will dramatically change the previous calculations for the cosmological radius and consequently everything after.
\par\noindent
Let us now employ the metric of a massless charged RNdS-like black hole.
First, upon combining Eq. \eqref{eq.12} and Eq. \eqref{eq.22}, the GUP-corrected entropy $\Delta S$ ends up being
\bea\label{eq.23}
\Delta S & =& -\int \left(\frac{\alpha^2}{\beta^2}-\frac{\alpha}{4\beta}\right) dA_c\\
& =& \int \left[- \frac{\alpha^2}{4\pi^2} \left(\frac{4\Lambda^2}{9}r^2_c - \frac{4\Lambda}{3}+\frac{1}{r_c^2}\right) 
\right.\nonumber\\
&&\left.+ \frac{\alpha}{8\pi}\left(\frac{2\Lambda}{3}r_c - \frac{1}{r_c}\right)\right]dA_c\\
& =& \left.- \frac{2\alpha^2}{\pi} \left(\frac{\Lambda^2}{9}r^4_c - \frac{2\Lambda}{3}r_c^2+\ln{r_c}\right)\right.\nonumber\\
&& + \alpha\left(\frac{2\Lambda}{9}r_c^3 - r_c\right)~.
\eea
Then, maximizing the entropy with respect to $\Lambda$ in the light of the extra $\alpha$-term in $\Delta S$, the corresponding cosmological radius is determined by both $\alpha$ and $\Lambda$ together according to the equation
\be 
 r_c^2 - \frac{\pi}{2\alpha\Lambda} r_c - \frac{3}{\Lambda}=0
 \ee
 which gives the root
 \be
r_c=\left(\frac{\pi}{4\alpha\Lambda}\right)+\sqrt{\left(\frac{\pi}{4\alpha\Lambda}\right)^2+\frac{3}{\Lambda}}~.
\label{eq.24}
\ee
It is noteworthy that the other root in order to be positive demands $\Lambda<0$, which of course 
contradicts the fact that spacetime is de Sitter.
\par\noindent
For the sake of comparison with the result obtained in the previous section, namely $r_c=\sqrt{3/\Lambda}$, we expand the root given in Eq. \eqref{eq.24}  up to $\mathcal{O}(\alpha^3)$ to obtain
\be\label{eq.25}
r_c \sim \left(\frac{\pi}{4\alpha\Lambda}\right) + \sqrt{\frac{3}{\Lambda}}\left[1+\frac{1}{2}\left(\frac{\pi}{4\alpha\Lambda}\right)^2\frac{\Lambda}{3}\right]~.
\ee
At this point a couple of comments are in order. First, when we are in strong gravity regimes, e.g. near black hole horizons, which  can be viewed as $\alpha \to \infty$, from Eq. \eqref{eq.25} we obtain $r_c\to\sqrt{3/\Lambda}$ which agrees with what was presented in the previous section and proven in Ref. \cite{Xiang:2005wg}. 
Second, if we employ Eq. \eqref{eq.24},  equation $f(r_c, Q, \Lambda)=0$ will be satisfied for  an electric charge $Q$ 
of the form
\be\label{eq.26}
|Q| = \frac{1}{8 \sqrt{6}} & \bigg(\frac{4 \pi ^4}{\alpha ^4 \Lambda ^3}+\frac{144 \pi ^2}{\alpha ^2 \Lambda ^2}\\
& + \frac{\pi  \left(48 \alpha ^2 \Lambda +\pi ^2\right)^{3/2}}{\alpha ^4 \Lambda ^3}\\
& +\frac{3 \pi ^3 \sqrt{48 \alpha ^2 \Lambda +\pi ^2}}{\alpha ^4 \Lambda ^3}\bigg)^{1/2} \neq 0~.
\ee
It is clear that if we take the limit  $\alpha\to\infty$, from Eq. \eqref{eq.11} the electric charge will be $Q\to 0$, as expected. 
\par\noindent
Furthermore, since the electric charge receives a  nonzero value, namely $|Q|\neq 0$, solving equation $f(r,Q,\Lambda)=0$  with $f(r,Q,\Lambda)$ as given by Eq. \eqref{metric}, we get the roots (radii of horizons)
\bea\label{eq.27}
 r_{\pm\pm}  & = &\pm\sqrt{\frac{3\pm\sqrt{9+12Q^2\Lambda}}{2\Lambda}}\\
r_{\pm\mp}   & = & \pm\sqrt{\frac{3\mp\sqrt{9+12Q^2\Lambda}}{2\Lambda}}
\eea
\par\noindent
Since the radii of the horizons have to be positive, i.e., $r < 0$, the negative roots are excluded completely. 
The root $r_{++}$ is the largest one so it is the cosmological horizon \cite{Hobson:2006se}. The root $r_{+-}$ has also to be positive and, thus,  it is required $3-\sqrt{9+12Q^2\Lambda}>0$.  However, this gives $Q^{2}<0$ which is impossible thus the root $r_{+-}$ is unphysical and is also removed. Therefore, there is no event horizon, and the singularity of the RNdS-like black hole is a naked singularity. 
%
%
%
%
%
%
%
%
%
%
\section{Effect of linear and Quadratic GUP on Energy distribution of massless charged particles in RN$\text{d}$S-like spacetime}
%
%
%
%
%
\par\noindent
Assuming that the curvature singularity can store only a few bits of information, one may say that the Hawking radiation for the black hole under study will consist of massless charged particles. Thus, it is useful to compute  the total energy density of these massless charged particles.

Following the analysis in Ref.  \cite{Li:2009bh}, we calculate the total energy density of the massless charged particles in a general spherically symmetric static spacetime in the context of linear and quadratic  GUP. The first quantity to be employed for this calculation, is the invariant volume element of the phase space. 
 In Ref.  \cite{Chang:2001bm}, the invariant volume element of a phase space in a $D$-dimensional spacetime  was computed  in the context of linear GUP, while in Ref. \cite{Ali:2011ap} the invariant volume element of a phase space in $D$-dimensional spacetime  was computed  in the context of linear and quadratic GUP. However, in the latter case, the invariant volume was computed  to $\mathcal{O}(\alpha)$ while here we would like to be more precise and, thus, go up to $\mathcal{O}(\alpha^2)$. Therefore, the invariant volume element of a phase space in the context of linear and quadratic GUP to $\mathcal{O}(\alpha^2)$ is given as \cite{Vagenas:2019wzd}
\be \label{eq.28}
\frac{ d\mathbf{x}^D d\mathbf{p}^D}{(2\pi)^{D} \,(1-\alpha p + (\frac{2\alpha^2}{D+1}+\frac{\alpha^2}{2}) p^2)^{(D+1)}}~.
\ee
\par\noindent
At the WKB level, the norm of massless particle momentum 3-space vector is
\be \label{eq.29}
p^2=p_i p^i = \frac{w^2}{f}
\ee
where $f = f(r,Q,\Lambda)$ which  is given by Eq. \eqref{metric}. 
Setting $D=3$, the total energy density for all frequencies reads
\be\label{eq.30}
\hspace{-2ex}\rho(f,\beta)\!=\!\gamma\!\! \int\limits^{\infty}_0 \frac{f^{2} ~ w^3}{2\pi^2 (f-\alpha \sqrt{f} w +
 \alpha^2 w^2)^4}  \frac{dw}{\left(e^{\beta w}\pm 1\right)}
\ee
 where $\gamma$ is the spin degeneracy,  the minus stands for the massless charged bosons and  the plus stands for massless charged fermions.
Upon considering the change of variable $x=\beta w/2\pi$ and $T(r)=1/(\beta \sqrt{f})$, where  $T(r)$ is the local temperature, Eq. \eqref{eq.30} becomes
\be\label{eq.31}
\hspace{-2ex}\rho(x,T)\!=8\pi^2\gamma T^{4}\!\! \int \limits^{\infty}_0 \frac{x^3}{(1-ax+a^2 x^2)^4}\frac{dx}{\left(e^{2\pi x}\pm 1\right)}
\ee
where $a=2\pi\alpha T$. This integral is not quite easy to solve. However, at least for the bosons and up to $\mathcal{O}(\alpha^3)$, it looks close to Hurwitz zeta function
\be\label{eq.32}
\zeta(n,u)=\frac{1}{\Gamma(n)} \int \limits^{\infty}_0 \frac{x^{n-1} e^{-ux}}{1-e^x}dx
\ee
where $a, n>0$ . In this case, the total density given by Eq. \eqref{eq.31} is indeed a convergent integral. 
Thus, using contour integral techniques, Eq. \eqref{eq.31}  can be calculated, However, we focus more on demonstrating  the effect of GUP on such distribution(s) and, thus, we provide Figs. (\ref{fig.1}) and  (\ref{fig.2}). For fixed $\alpha$ and $T(r)$, we assume $a$ to be small compared with $x$. 
%
%
%
%
\begin{figure}[h!]
  \includegraphics[width=0.5\textwidth]{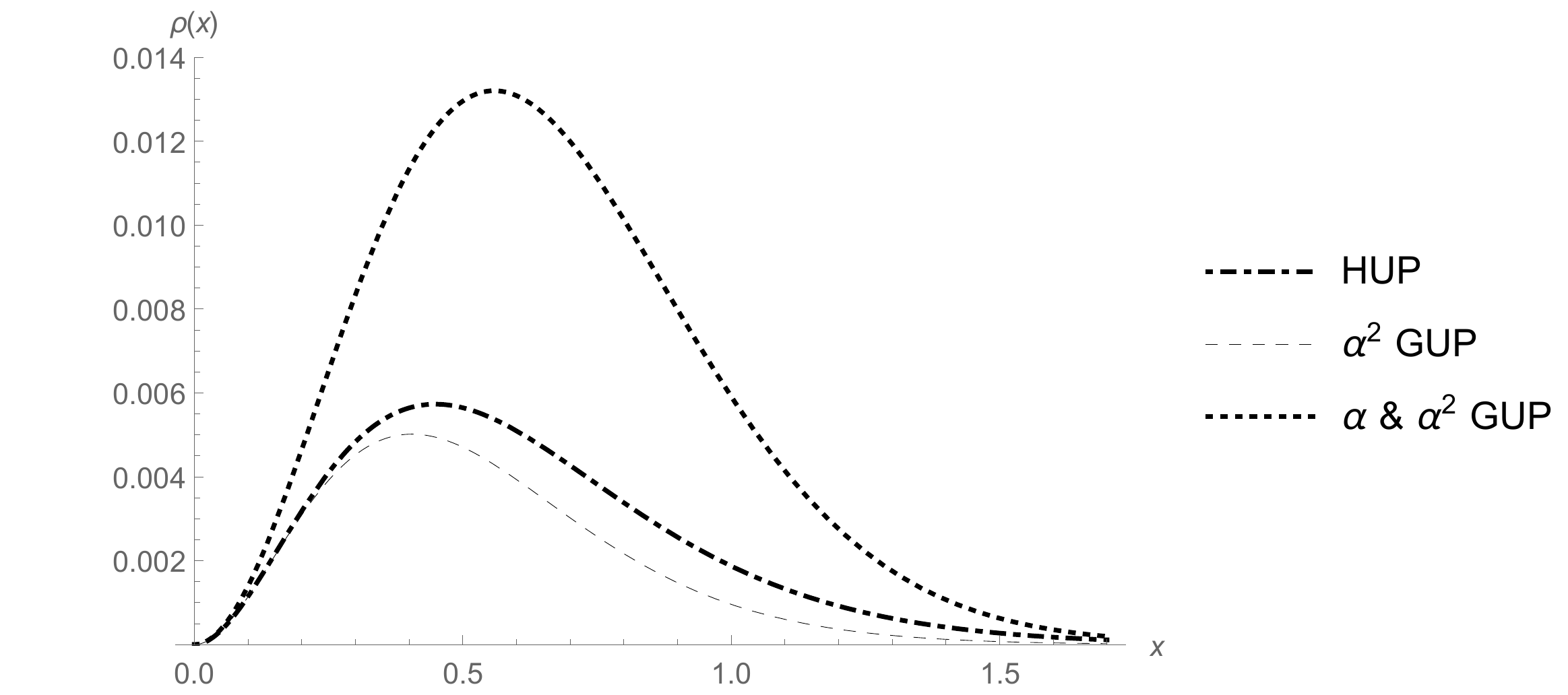}
  \caption{The total energy density $\rho(x)$ versus the variable $x$ for bosons with  $a=\frac{1}{2}$.}
  \label{fig.1}
\end{figure}
\begin{figure}[h!]
  \includegraphics[width=0.5\textwidth]{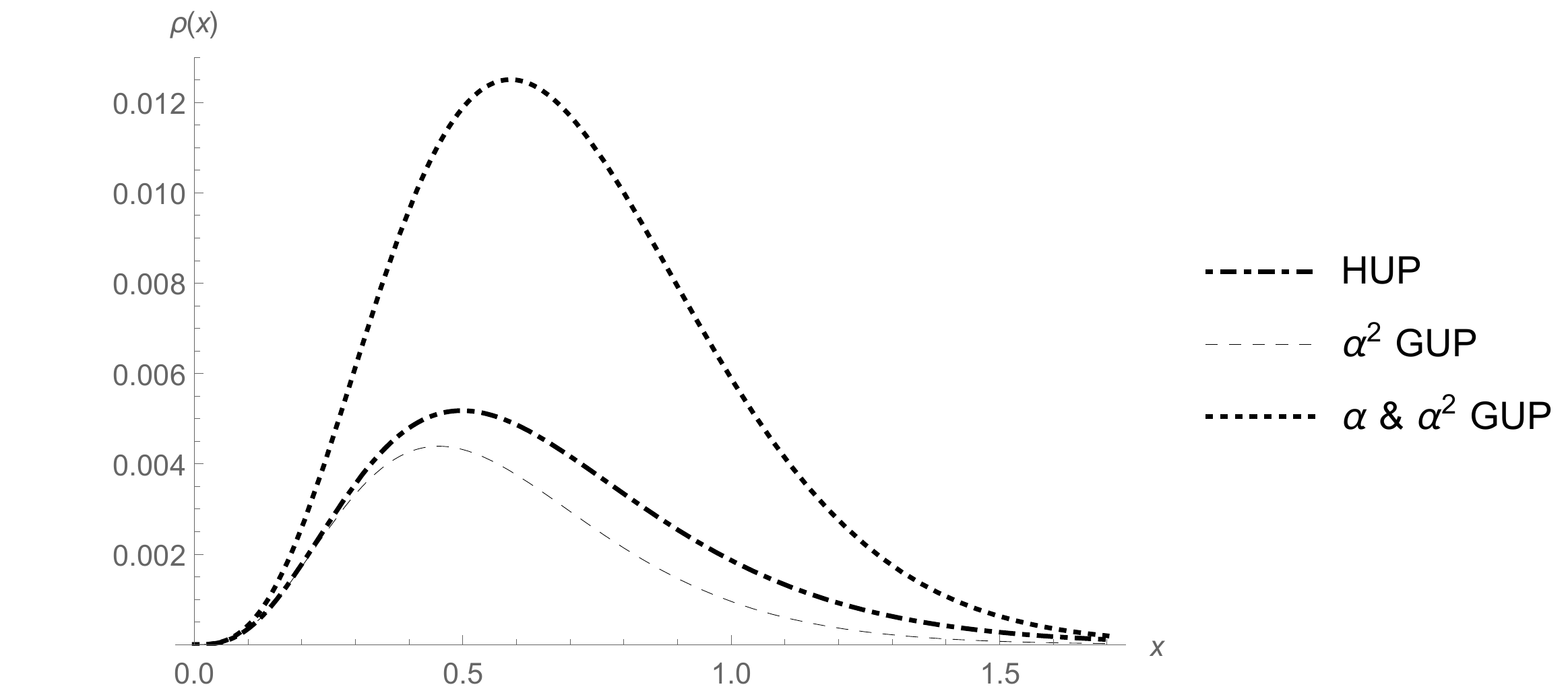}
  \caption{The total energy density $\rho(x)$ versus the variable $x$  for fermions with $a=\frac{1}{2}$.}  \label{fig.2}
\end{figure}
%
%
%
\begin{figure}[h!]
  \includegraphics[width=0.5\textwidth]{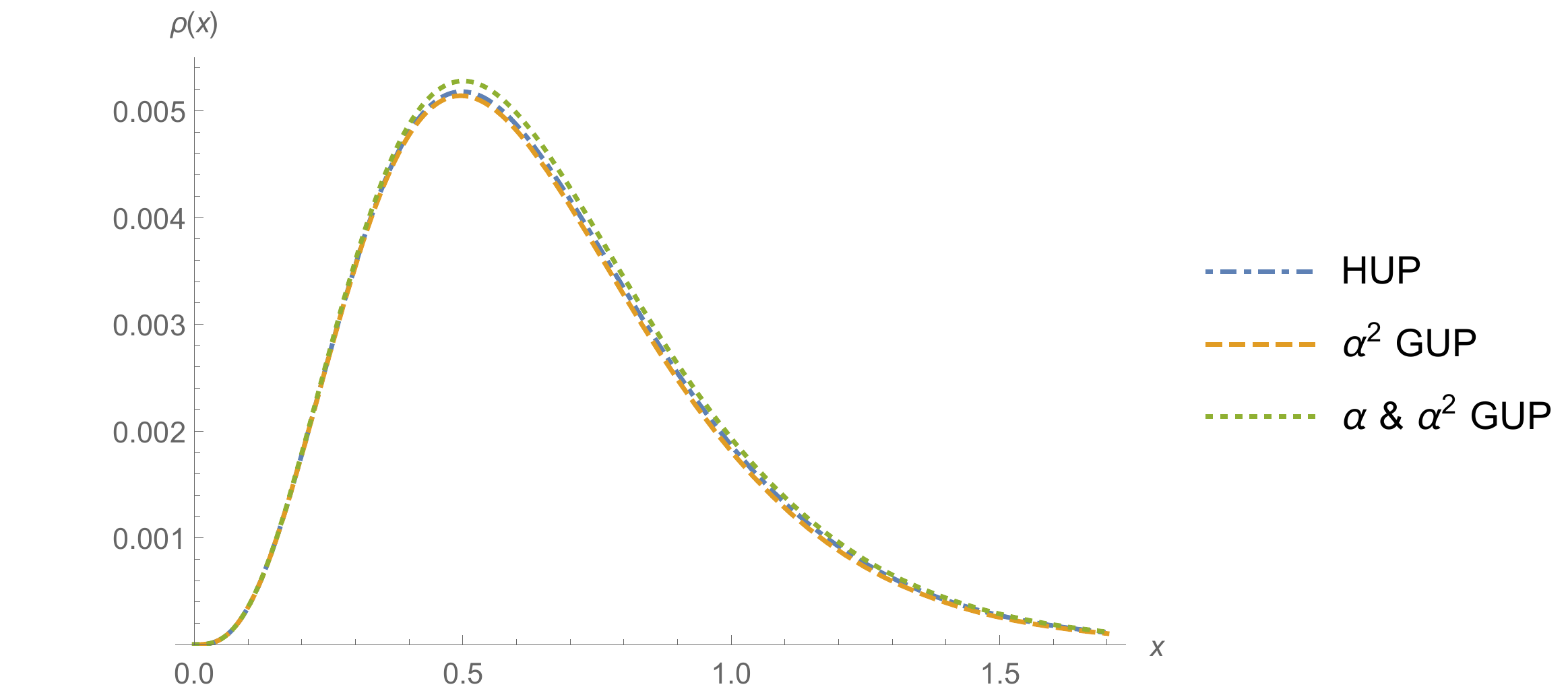}
  \caption{The total energy density  $\rho(x)$ versus the variable $x$  for bosons with $a=0.01$.}
  \label{fig.3}
\end{figure}
\begin{figure}[h!]
  \includegraphics[width=0.5\textwidth]{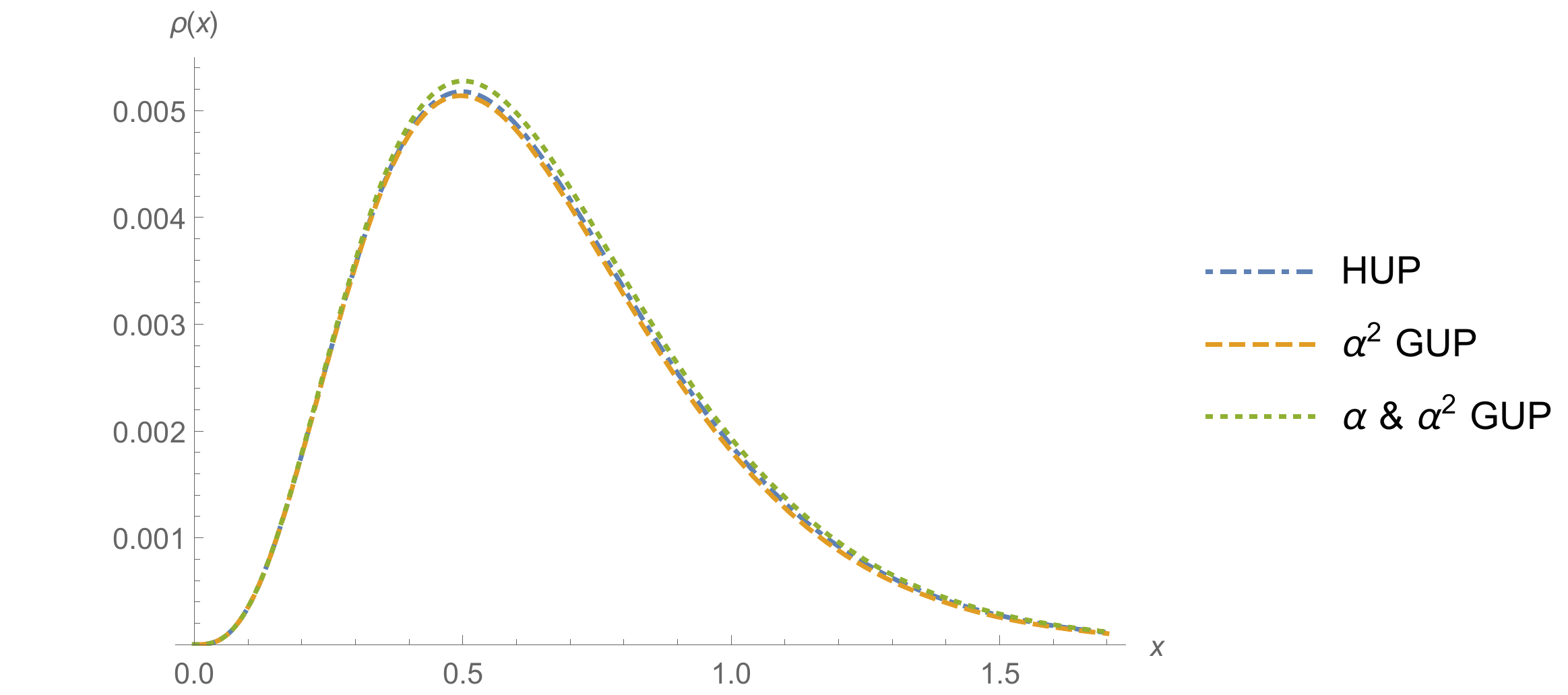}
  \caption{The total energy density  $\rho(x)$ versus the variable $x$  for fermions with $a=0.01$.}
  \label{fig.4}
\end{figure}
%
%
%
%
%
\par\noindent
A number of comments are in order here. 
First, for very diminutive values of $a$, as in Figs. (\ref{fig.3})  and (\ref{fig.4}),  we notice that the curves of GUP  tend to be that of HUP, as expected. 
Second, when such a collapsing system reaches the state of ultracold black hole, where $T(r) \sim \kappa_c \to 0$ \cite{Cai:1997ih}, and since the  constant $a$ is small only if $\alpha$ is also small,  it is evident that we have the GUP to tend to the HUP, as expected. 
Third, if we keep the GUP parameter $\alpha$ fixed and consider a different metric component $f \sim 1/T^2(r)$, say for instance the $f(M,r,Q,\Lambda)$ of the standard massive RNdS spacetime, to be compared with the larger $f(r,Q,\Lambda)$ of the massless RNdS spacetime for any allowed physical radius, then we notice that the massless one is colder, and hence $a$ is smaller for the massless RNdS spacetime, as expected. 
%
%
%
%
%
%
%
%
%
%
%
\section{Conclusions}
%
%
%
%
%
%
\par\noindent
In this work we have followed the methodology presented in Refs.  \cite{Li:2003bs,Xiang:2005wg} except we have introduced a  linear term in momentum in the GUP, namely linear and quadratic GUP, in order to investigate the GUP effect on a black hole system. We first computed the GUP-modified temperature and, using the first law of black hole mechanics, the GUP-modified entropy of the cosmological horizon. In our case, the entropy does not only have  a quadratic GUP correction term but also  a linear GUP correction term.  Then, for the specific massless RNdS spacetime, since the modified entropy of the cosmological horizon depends explicitly on the cosmological radius, we calculated the GUP-modified cosmological radius. Moreover, we expressed the electric charge of the specific black hole in terms of the GUP-modified cosmological horizon radius. Since the electric charge is nonzero, the equation for the locations of the black hole horizons is solved. The cosmological horizon is the only physical horizon, and thus it exists while there is no event horizon. Therefore, the singularity, i.e., $r=0$, is a naked singularity and, thus, the cosmic censorship conjecture is violated in the black hole spacetime under study. Furthermore, in Refs.  \cite{Li:2003bs,Xiang:2005wg}, it was also shown that by considering the quadratic GUP, the second law of black hole mechanics prevents the occurrence of massless charged particles. Assuming that the singularity can store only a few bits of information, one may say that the Hawking radiation for the black hole under study will consist of massless charged particles. 
For this reason, we also compute the total energy density of these massless charged particles in RNdS-like spacetime and in the presence of linear and quadratic GUP.
Our result does not say that massless charged particles can exist within the contemporary known Standard Model. Rather, contrary to \cite{Xiang:2005wg}, it says that until we get a concrete theory of quantum gravity, there is no physical principle that prohibits the existence of massless charged particles upon combining the second law of black hole mechanics together with the more general, linear and quadratic GUP. It is also worth noting that if massless charged particles  existed in low/moderate energies, they would have been detected easily.
\par\noindent
In contrast, this is not contradictory with the most famous hypothesis for massless charged particles where massless quarks are believed to exist at very high energies before symmetry breaking occurred (see for instance Ref. \cite{Fritzsch:2015jfa}). Despite we discuss an \emph{ultracold} black hole-like system, we showed that in the presence of linear and quadratic GUP, the corresponding massless charged particles have huge energy density compared with those of HUP and of the quadratic GUP. Even if those tentatively assumed massless charged particles are indeed not comparable to massless quarks in their features or the way of formation, the gigantic effect of gravitational collapse near the \emph{fundamental length}, that is necessary to form massless charged particles, might be comparable to the high energy condition to form massless quarks.
\par\noindent
Finally, we would like to emphasize on the subtlety of both fundamental topics, namely  the cosmic censorship conjecture and the massless charged particles, discussed here. It is our belief that the fundamental topics can not be resolved using phenomenological, semiclassical, and/or heuristic methodologies of quantum gravity \cite{Vagenas:2018pez}. Thus, we agree with Xiang and Shen on the indispensability of a full theory of quantum gravity theory to be applied when star collapses in order to get the required full picture of such phenomena.
%
%
%
%
%
%
%
 %
%
%
%
%
%
%
%
%
%
%
%
%
%
%
%

%
%
%
%
%
\end{arabicfootnotes}
\end{document}